\documentclass{aastex61}
\date{\today}
\begin{document}

\title{Median Statistics Estimate of the Distance to the Galactic Center}

\author{Tia Camarillo}
\altaffiliation{tiacamarillo@phys.ksu.edu}
\affiliation{Department of Physics, Kansas State University,116 Cardwell Hall, Manhattan, KS 66506, USA}

\author{Varun Mathur}
\altaffiliation{varunmathurv@gmail.com}
\affiliation{Department of Physics, Kansas State University,116 Cardwell Hall, Manhattan, KS 66506, USA}
\affiliation{Department of Physics, Birla Institute of Technology $\&$ Science, Pilani, Goa 403726, India}

\author{Tyler Mitchell}
\altaffiliation{tyler106@phys.ksu.edu}
\affiliation{Department of Physics, Kansas State University,116 Cardwell Hall, Manhattan, KS 66506, USA}

\author{Bharat Ratra}
\altaffiliation{ratra@phys.ksu.edu}
\affiliation{Department of Physics, Kansas State University,116 Cardwell Hall, Manhattan, KS 66506, USA}

\begin{abstract}
We show that error distributions of a compilation of 28 recent independent measurements of the distance from the Sun to the Galactic center, $R_{0}$, are wider than a standard Gaussian and best fit by an $n=4$ Student's $t$ probability density function. Given this non-Gaussianity, the results of our median statistics analysis, summarized as $R_{0}=8.0 \pm 0.3$ kpc ($2\sigma$ error), probably provides the most reliable estimate of $R_{0}$.
\end{abstract}


\section{INTRODUCTION}
The value of $R_{0}$, the distance of the Sun to the center of the Milky Way Galaxy, is a very important datum for astrophysics and cosmology. A quarter century ago, \cite{Reid1993} concluded that a reasonable summary value was $R_{0}=8.0 \pm 0.5$ kpc (errors are $1\sigma$ unless indicated otherwise). More recent summary estimates include $R_{0}=7.9 \pm 0.2$ kpc from \cite{Nikiforov2004}, $R_{0}=8.0 \pm 0.25$ kpc from \cite{Malkin2012}, $R_{0}=8.3 \pm 0.2\ \mathrm{(stat.)} \pm 0.4 \mathrm{\ (syst.)}$ kpc from \cite{deGrijs2016}, and $R_{0}=8.0 \pm 0.2$ kpc from \cite{Vallee2017}. 

De Grijs \& Bono (2016) compiled 273  $R_{0}$ measurements, not all of which are statistically independent, and carefully studied how publication bias might have influenced $R_{0}$ measurements. Their summary $R_{0}$ value is based on a consideration of only a very few of their 273 measurements. \cite{Vallee2017} on the other hand only compiled 27 very recent measurements, also not all independent; while we are able to reproduce his central estimate of $R_{0}=8.0$ kpc, we are unable to reproduce his $\pm 0.2$ kpc error bars from his compiled data set.

Here, we revisit the issue of determining a best estimate for, and errors on, $R_{0}$. Following \cite{Vallee2017}, we compile a list of 28 recent $R_{0}$ measurements in the belief that the more recent measurements are more reliable, but we carefully check to make sure that our list only includes statistically independent measurements, unlike the recent \cite{deGrijs2016} and \cite{Vallee2017} compilations.

Following, and generalizing, \cite{Chen2003}, we study the error distributions of this 28 measurement data set. We discover that the errors are somewhat non-Gaussian. This is not unexpected \citep{Bailey2017}; well-known examples of non-Gaussianity include Hubble constant $H_{0}$ measurements \citep{Chen2003}, $^{7}$Li abundance measurements \citep{Crandall2015, Zhang2017}, and LMC and SMC distance moduli measurements \citep{deGrijs2014, CrandallRatra2015}.

Significant effort is often devoted to determining whether there is intrinsic non-Gaussianity in astrophysical and cosmological systems \citep[e.g.,][]{Park2001, Ade2016}, as opposed to non-Gaussianity introduced by measurement techniques. This is because Gaussianity is assumed in many parameter constraint analyses \citep[e.g.,][]{Ratra1999, PodariuRatra2000}.

Care is required when analyzing data with non-Gaussian errors \citep[e.g.,][]{Gott2001, Bailey2017, Zhang2017}. \cite{Gott2001} developed median statistics partially for this purpose. Median statistics does not make use of the measurement errors and so is not affected by the non-Gaussianity, but since it discards some of the measurement information (the errors) it is less constraining. A well-known example of the use of median statistics is the analysis of $H_{0}$ measurements \citep{Gott2001, Chen2003, ChenRatra2011, Calabrese2012}.

In this paper, we apply median statistics to our compilation of 28 independent, recent $R_{0}$ measurements. We find $R_{0} = 7.96\ ^{+0.11}_{-0.23}$ ($^{+0.24}_{-0.30}$) kpc, where the errors are $1\sigma$ ($2\sigma$). For most practical purposes, this can be taken to be $R_{0}=8.0 \pm 0.3$ kpc at $2\sigma$.

In Sec. \ref{Data Compilation} we discuss our compilation of recent independent $R_{0}$ measurements and how it differs from that used by \cite{Vallee2017}. In Sec. \ref{Statistical Methods} we summarize our methods for computing central estimates and errors of the compiled data set. We outline five different error distributions in Sec. \ref{Error Distributions}. In Sec. \ref{Distribution Fitting} we present results from using the Kolmogorov-Smirnov test to match these error distributions to familiar functional forms, such as the Gaussian and Student's $t$, and tabulate the favored forms we find. We conclude in Sec. \ref{Conclusion}.

\section{DATA COMPILATION}
\label{Data Compilation}
The $R_{0}$ data we use in our analyses are listed in Table \ref{table:alldata}. The second column of the table lists the 27 $R_{0}$ values given to one decimal place in Table 1 of \cite{Vallee2017}. The third column of our Table \ref{table:alldata} updates these values, to two decimal places, from the original publications. 

Of these 27 measurements, only 20 are statistically independent, and these are listed in column 4 of Table \ref{table:alldata}. To these 20 measurements we added 8 new, post-2010, independent values that we found after a fairly exhaustive search of the literature. We decided to only use more recent (post-2010) data in the hope that they would be of better quality than earlier data. These 28 measurements are listed in column 5 of Table \ref{table:alldata}. Most of our analyses here focus on these 28 measurements.

In making our list of independent measurements, we ensure that no two estimates use the same experimental data. If two papers use the same method but use data from different equipment then we include both. Consider \cite{Boehle2016} and \cite{Gillessen2013}: both estimate $R_0$ by using the orbits of S-stars about the Galactic Center, Sgr $\mathrm{A}^*$. However, they use distinct experiments to constrain the orbits. There are quite a few papers that use the same method and data, from the same experiments, as the two above -- we include only the latest independent results and drop the rest. Some papers combine their result with other data: \cite{Do2013} combines their estimate of $R_0$ using statistical parallax with \cite{Ghez2008}, a predecessor of \cite{Boehle2016}. In this case we use the measurement of $R_0$ from \cite{Do2013}, that is not combined with \cite{Ghez2008} data, $R_0=8.92_{-0.55}^{+0.58}$ kpc. We assume that only a small degree of systematic error is present in measurements of $R_0$.\footnote{We do account for all stated systematic errors. Our results below, which show that the error distributions are not very non-Gaussian, are consistent with our assumption that unknown systematic errors are small.}

\begin{deluxetable*}{lccccr} 
\tablecaption{$R_{0}$ (in kpc) Measurements}
\tablewidth{0pt}
\tabletypesize{\scriptsize}
\tablehead{ 
\colhead{Year}& 
\colhead{Vall\'{e}e}&
\colhead{Vall\'{e}e:}& 
\colhead{Vall\'{e}e:}&
\colhead{Independent}&
\colhead{Reference}\\
\colhead{}& 
\colhead{}&
\colhead{updated\tablenotemark{a}}& 
\colhead{independent\tablenotemark{a}}&
\colhead{from 2011\tablenotemark{a}}&
\colhead{}
}
\startdata
2011  &    -    &    -    &    -    &  7.94  $\pm$  0.65  &  \cite{Fritz2011}\\
2011  &    -    &    -    &    -    &  8.07  $\pm$  0.35  &  \cite{Trippe2011}\\
2012  &  7.7  $\pm$  0.4  &  7.70  $\pm$  0.40  &    -    &    -    &  \cite{Morris2012}\\
2012  &  8.0  $\pm$  0.8  &  8.00  $\pm$  0.45  &  8.00  $\pm$  0.45  &  8.00  $\pm$  0.45  &  \cite{Bovy2012}\\
2012  &  8.0  $\pm$  0.4  &  8.05  $\pm$  0.45  &    -    &    -    &  \cite{Honma2012}\\
2012  &  8.3  $\pm$  0.4  &  8.27  $\pm$  0.29  &  8.27  $\pm$  0.29  &  8.27  $\pm$  0.29  &  \cite{Schonrich2012}\\
2013  &  7.6  $\pm$  0.6  &  7.50  $\pm$  0.60  &  7.50  $\pm$  0.60  &  7.50  $\pm$  0.60  &  \cite{Matsunaga2013}\\
2013  &    -    &    -    &    -    &  7.25  $\pm$  0.32  &  \cite{Bobylev2013}\\
2013  &  7.6  $\pm$  0.3  &  7.64  $\pm$  0.32  &  7.64  $\pm$  0.32  &  7.64  $\pm$  0.32  &  \cite{Bobylev2013}\\
2013  &    -    &    -    &    -    &  7.66  $\pm$  0.36  &  \cite{Bobylev2013}\\
2013  &    -    &    -    &    -    &  7.73  $\pm$  0.36  &  \cite{Dambis2013}\\
2013  &    -    &    -    &    -    &  7.91  $\pm$  0.41  &  \cite{Bono2013}\\
2013  &  8.0  $\pm$  0.8  &  7.98  $\pm$  0.79  &  7.98  $\pm$  0.79  &  7.98  $\pm$  0.79  &  \cite{Zhu2013}\\
2013  &  8.0  $\pm$  0.7  &  8.03  $\pm$  0.70  &  8.03  $\pm$  0.70  &  8.03  $\pm$  0.70  &  \cite{Zhu2013}\\
2013  &  8.2  $\pm$  0.8  &  8.25  $\pm$  0.79  &    -    &    -    &  \cite{Zhu2013}\\
2013  &  8.2  $\pm$  0.2  &  8.13  $\pm$  0.10\tablenotemark{b}  &  8.13  $\pm$  0.10\tablenotemark{b}  &  8.13  $\pm$  0.10\tablenotemark{b}  &  \cite{Cao2013}\\
2013  &  8.3  $\pm$  0.2  &  8.33  $\pm$  0.15  &    -    &    -    &  \cite{Dekany2013}\\
2013  &    -    &    -    &    -    &  8.20  $\pm$  0.34  &  \cite{Gillessen2013}\\
2013  &  8.5  $\pm$  0.4  &  8.46  $\pm$  0.40  &  8.92  $\pm$  0.57  &  8.92  $\pm$  0.57  &  \cite{Do2013}\\
2014  &  6.7  $\pm$  0.4  &  6.72  $\pm$  0.39  &  6.72  $\pm$  0.39  &  6.72  $\pm$  0.39  &  \cite{Branham2014}\\
2014  &  7.4  $\pm$  0.3  &  7.40  $\pm$  0.28  &  7.40  $\pm$  0.28  &  7.40  $\pm$  0.28  &  \cite{Francis2014}\\
2014  &  7.5  $\pm$  0.3  &  7.50  $\pm$  0.30  &  7.50  $\pm$  0.30  &  7.50  $\pm$  0.30  &  \cite{Francis2014}\\
2014  &  8.3  $\pm$  0.2  &  8.34  $\pm$  0.16  &    -    &    -    &  \cite{Reid2014}\\
2015  &    -    &    -    &    -    &  7.60  $\pm$  1.35  &  \cite{Ali2015}\\
2015  &  7.7  $\pm$  0.1  &  7.68  $\pm$  0.07  &  7.68  $\pm$  0.07  &  7.68  $\pm$  0.07  &  \cite{Branham2015}\\
2015  &  8.0  $\pm$  0.3  &  8.03  $\pm$  0.12  &    -    &    -    &  \cite{Bajkova2015}\\
2015  &  8.3  $\pm$  0.1  &  8.33  $\pm$  0.11  &  8.27  $\pm$  0.13  &  8.27  $\pm$  0.13  &  \cite{Chatzopoulos2015}\\
2015  &  8.3  $\pm$  0.4  &  8.27  $\pm$  0.40  &  8.27  $\pm$  0.40  &  8.27  $\pm$  0.40  &  \cite{Pietrukowicz2015}\\
2015  &  8.3  $\pm$  0.3  &  8.30  $\pm$  0.25  &  8.30  $\pm$  0.25  &  8.30  $\pm$  0.25  &  \cite{Kupper2015}\\
2016  &  7.9  $\pm$  0.1  &  7.86  $\pm$  0.15  &  7.86  $\pm$  0.15  &  7.86  $\pm$  0.15  &  \cite{Boehle2016}\\
2016  &  8.4  $\pm$  0.1  &  8.24  $\pm$  0.12  &  8.24  $\pm$  0.12  &  8.24  $\pm$  0.12  &  \cite{Rastorguev2016}\\
2016  &  8.9  $\pm$  0.4  &  8.90  $\pm$  0.40  &  8.90  $\pm$  0.40  &  8.90  $\pm$  0.40  &  \cite{Catchpole2016}\\
2017  &  7.6  $\pm$  0.1  &  7.64  $\pm$  0.09  &  7.64  $\pm$  0.09  &  7.64  $\pm$  0.09  &  \cite{Branham2017}\\
2017  &  8.0  $\pm$  0.2  &  7.97  $\pm$  0.15  &    -    &    -    &  \cite{McMillan2017}\\
2017  &  8.2  $\pm$  0.1  &  8.20  $\pm$  0.09  &  8.20  $\pm$  0.09  &  8.20  $\pm$  0.09  &  \cite{McMillan2017}\\
\noalign{\vskip 1mm}
\enddata
\tablenotetext{a}{We determine the error by symmetrizing the error bars (if necessary) and adding the statistical and systematic errors in quadrature.}
\tablenotetext{b}{\cite{Cao2013} does not list an error bar. We thank L. Cao and S. Mao for providing the value listed here via private communication (2017).}
\label{table:alldata}
\end{deluxetable*}

\section{SUMMARY OF METHODS}
\label{Methods}
To construct error distributions of our data sets, we use three central estimates: the median, weighted mean, and arithmetic mean.\footnote{We follow the conventions of Secs. 38 and 39 of \cite{PDG}.}

\subsection{Statistical Methods}
\label{Statistical Methods}
Median statistics benefits from ignoring the measurements' individual errors, at the expense of having a larger uncertainty about the median than that of a method that utilizes the error information. For a sufficiently large number of statistically-independent values, it is expected that there exists a true median with half of the data points lying above and below it. Each individual measurement has a $50\%$ probability of being greater or less than the true median. \cite{Gott2001} explains that for $i=1,2,....,N$ independent measurements ${M}_{i}$, the probability of the median falling between ${M}_{i}$ and ${M}_{i+1}$ follows the binomial distribution 
\begin{equation}
\label{Gott}
{P}={\frac{{{2}^{-N}}{N!}}{{i!}(N-1)!}}.
\end{equation}
The one (two) standard deviation error associated with the median is defined in \cite{Gott2001} as the range about the median including $68.27\%$ ($95.45\%$) of the probability.\footnote{For other discussions and applications of median statistics, see \cite{ChenRatra2003}, \cite{Mamajek2008}, \cite{Andreon2012}, \cite{Farooq2013}, \cite{Croft2015}, \cite{Ding2015}, \cite{Groener2016}, \cite{Zheng2016}, \cite{Farooq2017}, \cite{Leaf2017}, and \cite{Sereno2017}.} The one standard deviation error given by a \cite{Gott2001} $68.27\%$ confidence range is smaller than that obtained by binning the measurements and integrating outwards to $68.27\%$ of the total area around the median \citep{CrandallRatra2014}. We call the error determined from the probability distribution of eq. (\ref{Gott}) $\sigma_{\mathrm{Gott}}$, while we refer to the result from the integration of the binned measurements' method as $\sigma_{\mathrm{med}}$. 

Utilizing the idea that ``better" measurements should have more weight, weighted mean statistics yields the benefit of a smaller error about the central estimate and takes the risk of under-weighting values with inaccurate uncertainties \citep[see, e.g.,][]{Podariu2001}. The weighted mean is defined as
\begin{equation}
{{M}}_{\mathrm{wm}}=\frac{\sum\limits_{i=1}^{N}{{M}}_{i}/\sigma_{i}^2}{\sum\limits_{i=1}^{N}1/\sigma_{i}^2},
\end{equation}
where $\sigma_{i}$ are the one standard deviation errors. The weighted mean standard deviation is
\begin{equation}
\sigma_{\mathrm{wm}}=\frac{1}{\sqrt{\sum_{i=1}^{N}1/\sigma_{i}^{2}}}.
\end{equation}
In our weighted mean analysis, and other analyses that use the errors, $\sigma_{i}$ is the quadrature sum of the (symmetrized) statistical and systematic (if quoted) errors.

\begin{deluxetable*}{lcccc} 
\tablecolumns{5}
\tablecaption{$R_{0}$ (in kpc) Central Estimates and Errors}
\tablewidth{0pt}
\tabletypesize{\scriptsize}
\tablehead{ 
\colhead{}& 
\colhead{Vall\'{e}e}&
\colhead{Vall\'{e}e:}& 
\colhead{Vall\'{e}e:}&
\colhead{Independent}\\
\colhead{}& 
\colhead{}&
\colhead{updated}& 
\colhead{independent}&
\colhead{from 2011}
}
\startdata
Median, integral\tablenotemark{a} & $8.00\ ^{+0.36}_{-0.34}\ ^{+0.54}_{-1.26}$ & $8.03\ ^{+0.31}_{-0.32}\ ^{+0.83}_{-1.27}$ & $8.02\ ^{+0.26}_{-0.55}\ ^{+0.86}_{-1.24}$ & $7.96\ ^{+0.29}_{-0.50}\ ^{+0.90}_{-1.20}$\\
$1\sigma$ range & $7.66-8.36$ & $7.71-8.34$ & $7.47-8.28$ & $7.46-8.25$\\
$2\sigma$ range & $6.74-8.54$ & $6.76-8.86$ & $6.78-8.88$ & $6.76-8.86$\\\\
Median, Gott\tablenotemark{b} & $8.00\ ^{+0.20}_{-0.00}\ ^{+0.30}_{-0.30}$ & $8.03\ ^{+0.17}_{-0.05}\ ^{+0.24}_{-0.33}$ & $8.02\ ^{+0.18}_{-0.16}\ ^{+0.25}_{-0.38}$ & $7.96\ ^{+0.11}_{-0.23}\ ^{+0.24}_{-0.30}$\\
$1\sigma$ range & $8.00-8.20$ & $7.98-8.20$ & $7.86-8.20$ & $7.73-8.07$\\
$2\sigma$ range & $7.70-8.30$ & $7.70-8.27$ & $7.64-8.27$ & $7.66-8.20$\\\\
Weighted Mean & $8.02\pm0.04$ & $7.99\pm0.03$ & $7.93\pm0.03$ & $7.93\pm0.03$\\
$1\sigma$ range & $7.99-8.06$ & $7.95-8.02$ & $7.90-7.97$ & $7.89-7.96$\\\\
Arithmetic Mean & $8.00\pm0.08$ & $7.99\pm0.08$ & $7.97\pm0.11$ & $7.92\pm0.09$\\
$1\sigma$ range & $7.91-8.08$ & $7.91-8.07$ & $7.86-8.08$ & $7.84-8.01$\\
\noalign{\vskip 1mm}
\enddata
\tablenotetext{a}{Errors are estimated by binning the measurements to 0.1 kpc and integrating outwards until reaching 68.27\% and 95.45\% of the area under the distribution.}
\tablenotetext{b}{Errors are estimated from the median statistics probability distribution of eq. (\ref{Gott}).}
\label{table:CEs}
\end{deluxetable*}

It may also be of value to consider the arithmetic mean,
\begin{equation}
{{M}}_{\mathrm{m}}=\frac{1}{N}{\sum\limits_{i=1}^{N}{{M}}_{i}}.
\end{equation}
The underlying assumptions here are that each of the measurements have roughly the same uncertainty, and that the data come from a normally distributed set. The standard error of the mean is
\begin{equation}
\sigma_{\mathrm{m}}=\sqrt{{\frac{1}{N^2}}\sum_{i=1}^{N}({M}_{i} - {M}_{\mathrm{m}})^{2}} .
\end{equation}
Note that the standard deviation of the data set, $\sigma$, and the standard error of the mean, $\sigma_m$, differ by the square root of the amount of measurements: $\sigma_m = \sigma / \sqrt{N}$.

The central estimates and associated errors are recorded in Table \ref{table:CEs} for each of the data sets of Table \ref{table:alldata}. From column 2 of Table \ref{table:CEs}, we see our median, weighted mean, and arithmetic mean central estimates of 8 kpc coincide with those of \cite{Vallee2017} (at the bottom of his Table 1). However, we are unable to reproduce his weighted mean and arithmetic mean error bars of $\pm 0.2$ kpc (he does not quote a median error bar); our weighted (arithmetic) mean error bar is $\pm 0.04$ (0.4) kpc. 

The last column of Table \ref{table:CEs} summarizes our main result. As discussed below, we find the error distribution for our chosen 28 measurements are somewhat non-Gaussian, but not excessively so.\footnote{Seeing as the error distribution calculated from the median statistics of eq. (\ref{Gott}) is not very non-Gaussian it is unlikely that most errors have been incorrectly estimated. Specifically, it is unlikely that there are large undiscovered systematic errors.} Consequently we recommend that the median central value and the symmetrized errors for the 68.27\% and 95.45\% confidence ranges as defined in \cite{Gott2001} be used to describe the value of and errors on $R_{0}$. This gives $R_{0} = 7.96 \pm 0.17$ ($\pm 0.27$) kpc, with symmetrized $1\sigma$ ($2\sigma$) error, though it might be preferable to use the unsymmetrized result of $R_{0} = 7.96\ ^{+0.11}_{-0.23}$ ($^{+0.24}_{-0.30}$) kpc to take into account the slightly asymmetric nature of the set of measurements. For most practical purposes, $R_{0} = 8.0 \pm 0.3$ ($2\sigma$ error) serves as an appropriate summary estimate to one decimal place. 

\subsection{Error Distributions}
\label{Error Distributions}
After determining our central estimates, we construct our error distributions by using
\begin{equation}
\label{Nsig+}
N_{\sigma_{i}}=\frac{R_{i}-R_{\mathrm{CE}}}{\sqrt{\sigma_{i}^{2}+\sigma_{\mathrm{CE}}^{2}}}.
\end{equation}
Here $R_{\mathrm{CE}}$ is the central estimate of $R_{i}$ and $\sigma_{\mathrm{CE}}$ is the error of the central estimate of $R_{i}$. $N_{\sigma_{i}}$ represents how much $R_{i}$ deviates from the central estimate, taking into account both the error associated with the measurement and the error associated with the central estimate. In this paper we do not symmetrize $\sigma_{\mathrm{CE}}$ for the median statistics cases (the data are not symmetric enough to justify it). Thus, when applicable, we use the upper/right-side error $\sigma_{\mathrm{CE}}^{u}$ for when $R_{i} \geq R_{\mathrm{CE}}$ and the lower/left-side error $\sigma_{\mathrm{CE}}^{l}$ for when $R_{i} \leq R_{\mathrm{CE}}$. 

We label our error distributions $N^{\mathrm{med}}_{\sigma}$, $N^{\mathrm{Gott}}_{\sigma}$, $N^{\mathrm{wm+}}_{\sigma}$, and $N^{\mathrm{mean}}_{\sigma}$. These represent differing combinations of central estimates and errors, defined as
\begin{equation}
\label{NsigS}
N^{\mathrm{med}}_{\sigma_i}=\frac{R_{i}-R_{\mathrm{med}}}{\sqrt{\sigma_{i}^{2}+\sigma_{\mathrm{med}}^{2}}},
\end{equation}
\begin{equation}
\label{NsigG}
N^{\mathrm{Gott}}_{\sigma_i}=\frac{R_{i}-R_{\mathrm{med}}}{\sqrt{\sigma_{i}^{2}+\sigma_{\mathrm{Gott}}^{2}}},
\end{equation}
\begin{equation}
\label{NsigWM+}
N^{\mathrm{wm+}}_{\sigma_i}=\frac{R_{i}-R_{\mathrm{wm}}}{\sqrt{\sigma_{i}^{2}+\sigma_{\mathrm{wm}}^{2}}};
\end{equation}
\begin{equation}
\label{NsigM}
N^{\mathrm{mean}}_{\sigma_i}=\frac{R_{i}-R_{\mathrm{m}}}{\sqrt{\sigma_{i}^{2}+\sigma_{\mathrm{m}}^{2}}}.
\end{equation}
Since the central estimates are calculated from the data, they must to some degree be correlated with the error measurements. If the errors are Gaussian and the weighted mean has been determined from the measurements, then it is correlated with the measurements and a more appropriate error distribution is then\footnote{See the Appendix for a derivation.}
\begin{equation}
\label{NsigWM-}
N^{\mathrm{wm-}}_{\sigma_i}=\frac{R_{i}-R_{\mathrm{wm}}}{\sqrt{\sigma_{i}^{2}-\sigma_{\mathrm{wm}}^{2}}}.
\end{equation}
The derivation of an equivalent error distribution that accounts for the correlation is nontrivial for a median central estimate, however eq. (\ref{Nsig+}) provides a valuable limiting case.\footnote{It would be interesting to account for the correlation between the measurements and the median from eq. (\ref{Gott}), but this is beyond the scope of the current paper.}

We choose to use the above five error distributions to attempt to gain some insight into the $R_{0}$ measurements' error distribution.\footnote{We recognize that the integral method of calculating $\sigma_\mathrm{med}$ is not the error on the median itself (like the \cite{Gott2001} method provides) but is the deviation of the data set about the median. We include it to remain consistent with recently published results regarding the Gaussianity of error distributions where it was used in an attempt to also account for systematic uncertainties, e.g. \cite{CrandallRatra2014}. We propose for future analyses that this error not be regarded as the uncertainty on the median nor be used in calculating error distributions.}

\subsection{Distribution Fitting}
\label{Distribution Fitting}
We numerically study our error distributions using the one-sample Kolmogorov-Smirnov (K-S) test \citep{ModStatMeths}. This non-parametric, distribution-free test determines the probability that the given sample distribution comes from a well-defined probability density function (PDF), at a chosen significance level $\alpha$. In this paper we use Gaussian, Student's $t$, Cauchy, and Laplace (Double Exponential) distributions. The qualitative returns of a K-S test are a $D$ statistic and a $P$ value. The $D$ statistic is the supremum of, or the largest distance between, the cumulative sample distribution and the cumulative PDF. The closer this value is to zero, the better the sample distribution is well described by the PDF. For a sample distribution of $N$ measurements there is a critical value $D_{\mathrm{crit}}(N)$ that must be less than the test result, $D$, in order to not reject the null hypothesis at the specified significance level (which is conventionally set at $\alpha=0.05$ for a confidence level of $95\%$). For $N=28$ measurements $D_{\mathrm{crit}}=0.24993$.\footnote{See Appendix 3 of \cite{PRE} for a table of $D_{\mathrm{crit}}$ as a function of $N$.} The $P$ value follows from the $D$ statistic and represents not the probability that the sample set is from the proposed PDF, but rather the probability that we cannot reject the null hypothesis that the distributions are the same. It is for this reason that the probabilities of the K-S test should be used as qualitative indicators of distribution fitting. It is of interest to study the K-S test results for as many PDF's as possible. We choose the PDF with the lowest $D$ statistic and the highest $P$ value as the best representation of the error distribution under study.

\begin{deluxetable*}{@{\extracolsep{4pt}}llrlrlrlrlr} 
\tablecaption{K-S Test Probabilities \label{table:KS}}
\tablewidth{0pc}
\tabletypesize{\scriptsize}
\tablehead
{
\colhead{}& \multicolumn{2}{c}{$N^{\mathrm{med}}_{\sigma}$}& \multicolumn{2}{c}{$N^{\mathrm{Gott}}_{\sigma}$\  \tablenotemark{c}}& \multicolumn{2}{c}{$N^{\mathrm{wm+}}_{\sigma}$}& \multicolumn{2}{c}{$N^{\mathrm{wm-}}_{\sigma}$}& \multicolumn{2}{c}{$N^{\mathrm{mean}}_{\sigma}$}
\\ \cline{2-3} \cline{4-5} \cline{6-7} \cline{8-9} \cline{10-11}
\colhead{PDF}& \colhead{$S$\tablenotemark{a}}& \colhead{P$(\%)$\tablenotemark{b}}& \colhead{$S$\tablenotemark{a}}& \colhead{P$(\%)$\tablenotemark{b}}& \colhead{$S$\tablenotemark{a}}& \colhead{P$(\%)$\tablenotemark{b}}& \colhead{$S$\tablenotemark{a}}& \colhead{P$(\%)$\tablenotemark{b}}& \colhead{$S$\tablenotemark{a}}& \colhead{P$(\%)$\tablenotemark{b}}
}
\startdata
Gaussian & 1 & 69.4 & 1 & 53.4 & 1 & 11.9 & 1 & 11.7 & 1 & 17.8\\
Gaussian & 0.85 & 99.5 & 1.24 & 99.6 & 1.68 & 99.9 & 1.73 & 99.8 & 1.56 & 99.9\\
Laplace & 1 & 39.0 & 1 & 82.6 & 1 & 47.9 & 1 & 45.3 & 1 & 57.3\\
Laplace & 0.77 & 93.6 & 1.13 & 97.7 & 1.40 & 99.8 & 1.52 & 99.9 & 1.28 & 99.0\\
Cauchy & 1 & 4.1 & 1 & 32.8 & 1 & 64.6 & 1 & 88.7 & 1 & 50.2\\
Cauchy & 0.51 & 84.6 & 0.70 & 84.8 & 0.77 & 90.2 & 0.83 & 97.2 & 0.75 & 88.1\\
&\multicolumn{2}{c}{$n=100$}&\multicolumn{2}{c}{$n=3$}&\multicolumn{2}{c}{$n=2$}&\multicolumn{2}{c}{\dotfill\tablenotemark{e}}&\multicolumn{2}{c}{$n=2$}
\\ \cline{2-3} \cline{4-5} \cline{6-7} \cline{8-9} \cline{10-11}
Student's $t$\tablenotemark{d} & 1 & 67.7 & 1 & 97.5 & 1 & 81.1 & \dotfill & \dotfill & 1 & 88.8\\
&\multicolumn{2}{c}{$n=100$}&\multicolumn{2}{c}{$n=4$}&\multicolumn{2}{c}{$n=5$}&\multicolumn{2}{c}{$n=2$}&\multicolumn{2}{c}{$n=34$}
\\ \cline{2-3} \cline{4-5} \cline{6-7} \cline{8-9} \cline{10-11}
Student's $t$\tablenotemark{d}& 0.85 & 99.4 & 1.11 & 99.7 & 1.50 & 99.9 & 1.28 & 99.9 & 1.53 & 99.9\\
\noalign{\vskip 1mm}
\enddata
\tablenotetext{a}{Scale factor $S$ is first set at $S=1$ (representing the case when $|N_{\sigma}|=1$ corresponds to 1 standard deviation for a Gaussian distribution) and is then allowed to vary with the width of the function as $D$ is minimized.}
\tablenotetext{b}{This is the $P$ value described in Sec. \ref{Distribution Fitting}. It is the probability that we cannot reject the hypothesis that the sample distribution $N_{\sigma}$ came from a distribution created from the probability density function.}
\tablenotetext{c}{We use the errors corresponding to 68.27\% confidence in $N^{\mathrm{Gott}}_{\sigma}$ because we use 1 standard deviation for $N^{\mathrm{med}}_{\sigma}$.}
\tablenotetext{d}{We allow $n$ to vary between 1 and 100 for the Student's $t$ distribution.}
\tablenotetext{e}{The K-S test using a Student's $t$ PDF on $N^{\mathrm{wm-}}_{\sigma}$ for $S=1$ yielded a best fit of $n=1$ which is the Cauchy distribution.}
\end{deluxetable*}

We define our PDF's as functions of $|\textbf{N}|=|N_{\sigma}/S|$, where $S$ is a scale factor. When $S=1$ and $|\textbf{N}|=|N_{\sigma}|$, $P(|\textbf{N}|)$ is the standard form of the PDF. When $S>1$, the distribution is broader than the standard form, while $S<1$ corresponds to a narrower distribution. While $N_{\sigma_i}$ is computed with unsymmetrized errors, the distribution of $N_{\sigma}$ is symmetrized for the K-S test.

We define a Gaussian distribution of $N_{\sigma}$ with an expected $68.27\%$ and $95.45\%$ of the values falling within $|N_{\sigma}|\leq1$ and $|N_{\sigma}|\leq2$ respectively as
\begin{equation}
P(|\textbf{N}|)=\frac{1}{\sqrt{2\pi}}\exp(-|\textbf{N}|^{2}/2).
\end{equation}
The second distribution that we consider is a Laplace (Double Exponential), given by
\begin{equation}
P(|\textbf{N}|)=\frac{1}{2}\exp{(-|\textbf{N}|)}.
\end{equation}
The Laplace PDF is sharply peaked, with longer (smaller) tails than a Gaussian (Cauchy) distribution. For this distribution, $68.27\%$ and $95.45\%$ of the values correspond to $|N_{\sigma}|\leq1.2$ and $|N_{\sigma}|\leq3.1$ respectively. The Cauchy (Lorentz) distribution 
\begin{equation}
P(|\textbf{N}|)=\frac{1}{\pi}\frac{1}{1+|\textbf{N}|^{2}}
\end{equation}
has much higher probability in the tails, with an expected $68.27\%$ and $95.45\%$ of the values falling within $|N_{\sigma}|\leq1.8$ and $|N_{\sigma}|\leq14$ respectively. The Student's $t$ distribution is defined by
\begin{equation}
P(|\textbf{N}|)=\frac{\Gamma[(n+1)/2]}{\sqrt{\pi n}\ \Gamma(n/2)}\frac{1}{(1+|\textbf{N}|^{2}/n)^{(n+1)/2}}
\end{equation}
where $n$ is a positive non-zero parameter and $\Gamma$ is the gamma function. When $n=1$ this is the Cauchy distribution, and when $n\rightarrow\infty$ it becomes the Gaussian distribution. Thus, for $n>1$, it is a function with slightly less extended tails than a Cauchy, that decrease as $n$ increases. In this case, the limits corresponding to $68.27\%$ and $95.45\%$ of the values depend on the value of $n$.

Our K-S test results, for the 28 independent $R_{0}$ values listed in column 5 of Table \ref{table:alldata}, are shown in Table \ref{table:KS}. While some $S=1$ entries have low probabilities, and $P=11.7\%$ for the $S=1$ Gaussian case of the weighted mean central estimate and the $1\sigma$ error distribution of eq. (\ref{NsigWM-}), overall, allowing $S$ to vary a little away from unity, it is fair to conclude that the errors of the 28 measurement data set are not very non-Gaussian, although they are slightly so.\footnote{On the other hand, the corresponding analyses for the data sets of columns 2 and 3 of Table \ref{table:alldata} show that those 27 measurement data sets are more non-Gaussian, as might be expected, given the non-independence of some measurements.} Tables \ref{table:medNef} and \ref{table:medNlim}, which show the probabilities corresponding to $|N_{\sigma}|\leq1$ and $|N_{\sigma}|\leq2$ and the $|N_{\sigma}|$ values corresponding to 68.27\% and 95.45\% of the probability for these favored distributions, reinforce this conclusion. 

Columns 4 and 5 of Table \ref{table:KS} show the probabilities are as high as $99.9\%$ for a Gaussian distribution with $S=1.68$ and a Laplacian distribution with $S=1.52$, respectively. The non-Gaussianity associated with using the error bars from the $R_0$ measurements in weighted mean analyses can be substantiated from columns 4 and 5 of Tables \ref{table:medNef} and \ref{table:medNlim}: for the $S=1.68$ Gaussian in $N^{\mathrm{wm+}}_{\sigma}$, only $45\%$ ($77\%$) of the probability lies within $|N_{\sigma}|\leq1$ ($|N_{\sigma}|\leq2$) and to attain the standard probability of $68.27\%$ ($95.45\%$) we must integrate out to $|N_{\sigma}|=1.7$ ($|N_{\sigma}|=3.4$); for the $S=1.52$ Laplacian of $N^{\mathrm{wm-}}_{\sigma}$, only $48\%$ ($73\%$) of the probability lies within $|N_{\sigma}|\leq1$ ($|N_{\sigma}|\leq2$) and to attain the standard probability of $68.27\%$ ($95.45\%$) we must integrate out to $|N_{\sigma}|=1.7$ ($|N_{\sigma}|=4.7$). The Gaussian fits for $N^{\mathrm{wm+}}_{\sigma}$, $N^{\mathrm{wm-}}_{\sigma}$, and $N^{\mathrm{mean}}_{\sigma}$ require scale factors of $S=1.68$, $S=1.73$, and $S=1.56$ respectively. For this reason, it is best to use median statistics to determine the error bars on $R_0$, which are looser than those from weighted mean statistics and arithmetic mean statistics. The probability distribution computed from eq. (\ref{Gott}) then provides the best central estimate and errors bars for determining the somewhat non-Gaussian nature of the error distribution of the 28 independent $R_0$ measurements. The corresponding median-statistics error distribution of eq. (\ref{NsigG}) is best fit by an $n=4$ Student's $t$ PDF with an $S=1.1$ scale factor, and is non-Gaussian to the degree that with a probability of $99.6\%$, we cannot reject the hypothesis that it comes from a Gaussian distribution with an $S=1.24$ scale. The slightly broader-than-expected Gaussian distributed error distribution could indicate some (slightly) improperly estimated systematic uncertainties. This is, however, perhaps a mild concern until we can compile and study a larger set of recent and statistically independent measurements of $R_0$.

\section{CONCLUSION}
\label{Conclusion}
For more than three decades, the International Astronomical Union has recommended $R_{0} = 8.5$ kpc. In the last decade, evidence has been mounting that this might be a little too large \citep{Nikiforov2004, Malkin2012, deGrijs2016, Vallee2017}.

We have compiled a list of 28 recent, independent $R_{0}$ measurements. We find that the corresponding error distributions are slightly wider than a standard Gaussian. Consequently we believe a median statistics \citep{Gott2001} analysis provides a more reliable estimate of $R_{0}$ from this compilation. For most purposes $R_{0} = 8.0 \pm 0.3$ kpc ($2\sigma$ error), somewhat smaller than the 8.5 kpc IAU recommendation, is a reasonable summary of our results.

\acknowledgments
We thank D. Bailey, T. Bolton, S. Crandall, D. Pearson, J. Ryan, and L. Samushia for valuable conversations and recommendations. We also thank the referee, Gang Chen, for valuable comments. This work was supported in part by DOE grant DE-SC0011840, and with funding from an REU site funded by the National Science Foundation (NSF) and the Air Force Office of Scientific Research through NSF grant number PHYS-1461251.

\begin{deluxetable*}{@{\extracolsep{4pt}}llcclcclcclcclccr} 
\setlength{\tabcolsep}{4pt}
\tablecaption{$|N_{\sigma}|$ Expected Fractions}
\tabletypesize{\scriptsize}
\rotate
\tablewidth{0pt}
\tablehead
{
\colhead{}& \multicolumn{3}{c}{$N^{\mathrm{med}}_{\sigma}$}& \multicolumn{3}{c}{$N^{\mathrm{Gott}}_{\sigma}$}& \multicolumn{3}{c}{$N^{\mathrm{wm+}}_{\sigma}$}& \multicolumn{3}{c}{$N^{\mathrm{wm-}}_{\sigma}$}& \multicolumn{3}{c}{$N^{\mathrm{mean}}_{\sigma}$}\\
\cline{2-4} \cline{5-7} \cline{8-10}\cline{11-13} \cline{14-16}
\colhead{PDF}& \colhead{$S$\tablenotemark{a}}& \colhead{{$|N_{\sigma}|\leq1$}\tablenotemark{b}}& \colhead{{$|N_{\sigma}|\leq2$}\tablenotemark{b}}& \colhead{$S$\tablenotemark{a}}& \colhead{{$|N_{\sigma}|\leq1$}\tablenotemark{b}}& \colhead{{$|N_{\sigma}|\leq2$}\tablenotemark{b}}& \colhead{$S$\tablenotemark{a}}& \colhead{{$|N_{\sigma}|\leq1$}\tablenotemark{b}}& \colhead{{$|N_{\sigma}|\leq2$}\tablenotemark{b}}& \colhead{$S$\tablenotemark{a}}& \colhead{{$|N_{\sigma}|\leq1$}\tablenotemark{b}}& \colhead{{$|N_{\sigma}|\leq2$}\tablenotemark{b}}& \colhead{$S$\tablenotemark{a}}& \colhead{{$|N_{\sigma}|\leq1$}\tablenotemark{b}}& \colhead{{$|N_{\sigma}|\leq2$}\tablenotemark{b}}
}
\startdata
Gaussian&1&0.68&0.95&1&0.68&0.95&1&0.68&0.95&1&0.68&0.95&1&0.68&0.95\\
Gaussian&0.85&0.76&0.98&1.24&0.58&0.89&1.68&0.45&0.77&1.73&0.44&0.75&1.56&0.48&0.80\\
Laplace&1&0.63&0.87&1&0.63&0.87&1&0.63&0.87&1&0.63&0.87&1&0.63&0.87\\
Laplace&0.78&0.73&0.92&1.13&0.59&0.83&1.40&0.51&0.76&1.52&0.48&0.73&1.28&0.54&0.79\\
Cauchy&1&0.50&0.71&1&0.50&0.71&1&0.50&0.71&1&0.50&0.71&1&0.50&0.71\\
Cauchy&0.51&0.70&0.84&0.70&0.61&0.79&0.77&0.58&0.77&0.83&0.56&0.75&0.75&0.59&0.77\\\\
&\multicolumn{3}{c}{$n=100$}&\multicolumn{3}{c}{$n=3$}&\multicolumn{3}{c}{$n=2$}&\multicolumn{3}{c}{\dotfill\tablenotemark{c}}&\multicolumn{3}{c}{$n=2$}\\
\cline{2-4} \cline{5-7} \cline{8-10} \cline{11-13} \cline{14-16}
Student's $t$\dotfill&1&0.58&0.82& 1&0.61&0.86&1&0.58&0.82&\dotfill&\dotfill&\dotfill&1&0.58&0.82\\\\
&\multicolumn{3}{c}{$n=100$}&\multicolumn{3}{c}{$n=4$}&\multicolumn{3}{c}{$n=5$}&\multicolumn{3}{c}{$n=2$}&\multicolumn{3}{c}{$n=34$}\\
\cline{2-4} \cline{5-7} \cline{8-10} \cline{11-13} \cline{14-16}
Student's $t$\dotfill&0.85&0.76&0.98& 1.11&0.58&0.85&1.50&0.47&0.76& 1.28&0.48&0.74&1.53&0.48&0.80&\\\\
\noalign{\vskip -1mm}\hline\hline\noalign{\vskip 1mm}
Observed&\dotfill&0.86&1.00&\dotfill&0.54&0.93&\dotfill&0.50&0.71&\dotfill&0.50&0.68&\dotfill&0.50&0.71\\
\noalign{\vskip 1mm}
\enddata
\tablenotetext{a}{Scale factor $S$ is first set at $S=1$ (representing the case when $|N_{\sigma}|=1$ corresponds to 1 standard deviation for a Gaussian distribution) and is then allowed to vary with the width of the function as $D$ is minimized.}
\tablenotetext{b}{The fraction of data points that lie within $|N_{\sigma}|\leq1$ or $|N_{\sigma}|\leq2$.}
\tablenotetext{c}{The Student's $t$ test on $N_{\sigma_{\mathrm{wm-}}}$ for $S=1$ yielded a best fit of $n=1$ which is the Cauchy distribution.}
\label{table:medNef}
\end{deluxetable*}

\begin{deluxetable*}{@{\extracolsep{4pt}}llcclcclcclcclccr} 
\tablecaption{$|N_{\sigma}|$ Limits}
\tabletypesize{\scriptsize}
\rotate
\tablewidth{0pt}
\tablehead
{
\colhead{}& \multicolumn{3}{c}{$N^{\mathrm{med}}_{\sigma}$}& \multicolumn{3}{c}{$N^{\mathrm{Gott}}_{\sigma}$}& \multicolumn{3}{c}{$N^{\mathrm{wm+}}_{\sigma}$}& \multicolumn{3}{c}{$N^{\mathrm{wm-}}_{\sigma}$}& \multicolumn{3}{c}{$N^{\mathrm{mean}}_{\sigma}$}\\
\cline{2-4} \cline{5-7} \cline{8-10}\cline{11-13} \cline{14-16}
\colhead{PDF}& \colhead{$S$\tablenotemark{a}}& \colhead{{68.27\%}\tablenotemark{b}}& \colhead{{95.45\%}\tablenotemark{b}}& \colhead{$S$\tablenotemark{a}}& \colhead{{68.27\%}\tablenotemark{b}}& \colhead{{95.45\%}\tablenotemark{b}}& \colhead{$S$\tablenotemark{a}}& \colhead{{68.27\%}\tablenotemark{b}}& \colhead{{95.45\%}\tablenotemark{b}}& \colhead{$S$\tablenotemark{a}}& \colhead{{68.27\%}\tablenotemark{b}}& \colhead{{95.45\%}\tablenotemark{b}}& \colhead{$S$\tablenotemark{a}}& \colhead{{68.27\%}\tablenotemark{b}}& \colhead{{95.45\%}\tablenotemark{b}}
}
\startdata
Gaussian&1&1.0&2.0&1&1.0&2.0&1&1.0&2.0&1&1.0&2.0&1&1.0&2.0\\
Gaussian&0.85&0.9&1.7& 1.24&1.2&2.5& 1.68&1.7&3.4& 1.73&1.7&3.5& 1.56&1.6&3.1\\
Laplace&1&1.2&3.1&1&1.2&3.1&1&1.2&3.1&1&1.2&3.1&1&1.2&3.1\\
Laplace&0.78&0.9&2.4& 1.13&1.3&3.5& 1.40&1.6&4.3& 1.52&1.7&4.7& 1.28&1.5&4.0\\
Cauchy&1&1.8&14.0&1&1.8&14.0&1&1.8&14.0&1&1.8&14.0&1&1.8&14.0\\
Cauchy&0.51&0.9&7.0& 0.70&1.3&9.7& 0.77&1.4&10.6& 0.83&1.5&11.5& 0.75&1.4&10.6\\\\
&\multicolumn{3}{c}{$n=100$}&\multicolumn{3}{c}{$n=3$}&\multicolumn{3}{c}{$n=2$}&\multicolumn{3}{c}{\dotfill\tablenotemark{c}}&\multicolumn{3}{c}{$n=2$}\\
\cline{2-4} \cline{5-7} \cline{8-10} \cline{11-13} \cline{14-16}
Student's $t$\dotfill&1&1.0&2.0& 1&1.2&3.3& 1&1.3&4.5&\dotfill&\dotfill&\dotfill& 1&1.3&4.5\\\\
&\multicolumn{3}{c}{$n=100$}&\multicolumn{3}{c}{$n=4$}&\multicolumn{3}{c}{$n=5$}&\multicolumn{3}{c}{$n=2$}&\multicolumn{3}{c}{$n=34$}\\
\cline{2-4} \cline{5-7} \cline{8-10} \cline{11-13} \cline{14-16}
Student's $t$\dotfill& 0.85&0.9&1.7& 1.11&1.3&3.2& 1.50&1.7&4.0& 1.28&1.7&5.8& 1.53&1.5&3.2\\\\
\noalign{\vskip -1mm}\hline\hline\noalign{\vskip 1mm}
Observed&\dotfill&0.8&1.9&\dotfill&1.3&2.3&\dotfill&1.9&3.1&\dotfill&2.1&3.5&\dotfill&1.7&2.5\\
\noalign{\vskip 1mm}
\enddata
\tablenotetext{a}{Scale factor $S$ is first set at $S=1$ (representing the case when $|N_{\sigma}|=1$ corresponds to 1 standard deviation for a Gaussian distribution) and is then allowed to vary with the width of the function as $D$ is minimized.}
\tablenotetext{b}{The $|N_{\sigma}|$ limits containing 68.27\% or 95.45\% of the probability. For a Gaussian PDF with $S=1$, 68.27\% (95.45\%) of the probability is contained within $|N_{\sigma}|=1$ ($|N_{\sigma}|=2$).}
\tablenotetext{c}{The Student's $t$ test on $N_{\sigma_{\mathrm{wm-}}}$ for $S=1$ yielded a best fit of $n=1$ which is the Cauchy distribution.}
\label{table:medNlim}
\end{deluxetable*}

\appendix

\section{Derivation of eq. (11)} \label{Derivation}
While eq. (\ref{NsigWM-}) is well known to practitioners, we have been unable to find a derivation of it, and so provide this here.

For $i=1,2,..,N$ measurements $M_i$ with individual errors $\sigma_i$, modeled to be Gaussian about a central estimate with $M_\mathrm{CE}$ which itself has uncertainty $\sigma_\mathrm{CE}$, we define an uncertainty-normalized difference
\begin{equation}
N_{\sigma_{i}}=\frac{M_{i}-M_{\mathrm{CE}}}{\sqrt{\sigma_{i}^{2}+\sigma_{\mathrm{CE}}^{2}}}.
\end{equation}
This is the number of standard deviations a particular measurement differs from the central value. If we use a central estimate like the weighted mean, we can again standardize an $N^\mathrm{wm}_\sigma$. We begin by defining the weighted mean and it's error:
\begin{equation}
M_{\mathrm{wm}}=\frac{\sum_{i=1}^{N} M_{i} / \sigma_{i}^{2}}{\sum_{i=1}^{N} 1 / \sigma_{i}^{2}}
\end{equation}
and \citep{Podariu2001}
\begin{equation}
{\frac{1}{\sigma_{\mathrm{wm}}^2}}=\sum_{i=1}^{N}\frac{1}{\sigma_{i}^2}.
\end{equation}

However, a problem arises depending on how correlated $M_i$ and $M_\mathrm{CE}$ are. Defining $D_i$ that can be normalized to find a standardized $N_\sigma$ where
\begin{equation}
{D_i}=M_i - M_{\mathrm{wm}},
\end{equation}
we can calculate the variance of this quantity to later use for normalization
\begin{equation}
{\mathrm{Var}(D_i)}=\mathrm{Var}(M_i - M_{\mathrm{wm}}).
\end{equation}
If $M_i$ and $M_{\mathrm{wm}}$ are independent, the variance is distributed as
\begin{equation} \label{variance}
{\mathrm{Var}(aX+bY)}=a^2 \mathrm{Var}(X) + b^2 \mathrm{Var}(Y)
\end{equation}
and it is this case that yields the well-known result of adding errors in quadrature. As they are correlated though, let's try a different approach. The variance becomes
\begin{equation}
{\mathrm{Var}(D_i)}=\mathrm{Var}\left(M_i - \frac{\sum_{j=1}^{N} M_{j} / \sigma_{j}^{2}}{\sum_{k=1}^{N} 1 / \sigma_{k}^{2}}\right)
\end{equation}
which can be rearranged as
\begin{equation}
{\mathrm{Var}(D_i)}=\mathrm{Var}\left[\left(1- \frac{1 / \sigma_{i}^{2}}{\sum_{k=1}^{N} 1 / \sigma_{k}^{2}}\right)M_i - \frac{\sum_{j\neq i}^{N} M_{j} / \sigma_{j}^{2}}{\sum_{k=1}^{N} 1 / \sigma_{k}^{2}}\right].
\end{equation}
Here, we make the assumption that the measurements were made independently. Using eq. (\ref{variance}), the above becomes
\begin{equation}
{\mathrm{Var}(D_i)}=\left(1- \frac{1 / \sigma_{i}^{2}}{\sum_{k=1}^{N} 1 / \sigma_{k}^{2}}\right)^2 \mathrm{Var}(M_i) + \frac{\sum_{j\neq i}^{N} \mathrm{Var}(M_j) / \sigma_{j}^{4}}{(\sum_{k=1}^{N} 1 / \sigma_{k}^{2})^2}
\end{equation}
which can be simplified by opening the squares and by sending $\mathrm{Var}(M_i)$ into the summation over $N$
\begin{equation} \label{opened square}
{\mathrm{Var}(D_i)}=\left(1- 2\frac{1 / \sigma_{i}^{2}}{\sum_{k=1}^{N} 1 / \sigma_{k}^{2}}\right) \mathrm{Var}(M_i) + \frac{\sum_{j=1}^{N} \mathrm{Var}(M_j) / \sigma_{j}^{4}}{(\sum_{k=1}^{N} 1 / \sigma_{k}^{2})^2}.
\end{equation}
Now we make the assumption that the $M_i$ are Gaussianly distributed with variance $\sigma_i^2$, an assumption made even in the case of adding errors in quadrature, as in \cite{Bailey2017}. It follows then that
\begin{equation} 
{\mathrm{Var}(D_i)}=\left(1- 2\frac{1 / \sigma_{i}^{2}}{\sum_{k=1}^{N} 1 / \sigma_{k}^{2}}\right) \sigma_i^2 + \frac{\sum_{j=1}^{N}1 / \sigma_{j}^{2}}{(\sum_{k=1}^{N} 1 / \sigma_{k}^{2})^2}  =\sigma_i^2 - \sigma_{\mathrm{wm}}^{2}
\end{equation}
This gives the new equation that is better suited for correlated values,
\begin{equation}
N_{\sigma_{i}}=\frac{D_i}{\sqrt{\mathrm{Var}(D_i)}}=\frac{M_{i} - M_\mathrm{wm}}{\sqrt{\sigma_{i}^{2} - \sigma_{\mathrm{wm}}^{2}}}
\end{equation}
which may look familiar to some as the pull of a Gaussian measurement $M_i$ from the average value $M_\mathrm{CE}$ determined from the set of measurements. 

It should be noted that the median and arithmetic mean determined from the measurements are also correlated with the data and in a more careful analysis this should be accounted for. It may be possible to account for the median's correlation to the data using a Monte Carlo analysis (this requires knowledge of the data distribution which depends on the central estimate in question). We hope to discuss this elsewhere.

\end{document}